\documentclass[12pt]{article}

\usepackage{scicite}
\usepackage{graphicx}
\usepackage[hyphens]{url}
\usepackage{times}

\newenvironment{sciabstract}{%
\begin{quote} \bf}
{\end{quote}}

\title{Propagation of the Madden--Julian oscillation as a deterministic chaotic phenomenon}

\author
{Daisuke Takasuka,$^{1\ast}$ Tamaki Suematsu,$^{2}$ Hiroaki Miura$^{3}$, Masuo Nakano$^{4, 5}$\\
\\
\normalsize{$^{1}$Department of Geophysics, Tohoku University,}\\
\normalsize{6-3 Aramaki-aza-aoba, Aoba-ku, Sendai 980-8578, Japan}\\
\normalsize{$^{2}$RIKEN Center for Computational Science}\\
\normalsize{$^{3}$Department of Earth and Planetary Science, The University of Tokyo}\\
\normalsize{$^{4}$Japan Agency for Marine-Earth Science and Technology}\\
\normalsize{$^{5}$Typhoon Science and Technology Research Center, Yokohama National University}\\
\normalsize{$^\ast$E-mail:  takasuka@tohoku.ac.jp}
}

\date{}

\begin{document} 


\baselineskip24pt


\maketitle

\newpage
\begin{sciabstract}
The Madden--Julian oscillation (MJO), a gigantic tropical weather system, is marked by 
eastward travel of cumulus cloud clusters over the Indo-Pacific region and often 
causes severe weather and climate events worldwide. The physics and predictability 
of MJO propagation remain elusive, partly because of little attention to untangling roles 
of multi-scale processes relevant to the MJO. Here, we reveal the chaotic nature of 
MJO propagation arising from cross-scale nonlinear interactions, based on 4,000-member 
ensemble global cloud-system-resolving simulations of two MJO events. 
Against conventional linearized thinking, multiple regimes with distinct timings of 
MJO propagation emerge under a single atmosphere-ocean background. The bifurcation 
emergence depends critically on the equatorial asymmetry of climatological sea surface 
temperature. Selection of the bifurcated regimes is probabilistic, influenced by whether 
tropical-extratropical interplay promotes moistening associated with westward-propagating 
tropical waves over the western Pacific. These aspects help build a comprehensive MJO 
model and foresee when the MJO propagates.

\end{sciabstract}
\newpage


\section*{Introduction}
The Madden--Julian oscillation (MJO) is the most predominant variability in the tropical 
atmosphere, exerting far-reaching impacts on global climate and weather patterns 
through the substantial subseasonal-scale modulation of equatorial rainfall and 
large-scale wind fields \cite{Zhang2013a}. It manifests as an $O(10^3)$-km 
scale cluster of clouds traveling eastward over the Indo-Pacific region at a speed of 
around 5 m s$^{-1}$ \cite{Madden1972}. It is called the ``storm king" \cite{Hand2015} 
and described as ``{\it the last type of weather system for which the basic physical 
mechanisms are not well understood}" \cite{Randall2012}. 
MJO propagation into the tropical western Pacific (WP) has been known to 
trigger the phase transition of the El Ni\~{n}o/Southern Oscillation 
\cite{McPhaden1999, Takayabu1999, Hendon2007}, awake the Asian/Australian 
monsoon \cite{Yasunari1979, Hendon1990b}, influence tropical cyclogenesis 
\cite{Maloney2000}, and even affect mid-latitude weather extremes [e.g., heat/cold 
waves, heavy rainfall, and tornados \cite{Cerne2011, Jeong2005, Jones2012, 
Thompson2013}] as well as stratospheric temperature and circulation in the Arctic 
\cite{Liu2014, Garfinkel2012}. With these connections between weather and climate 
systems via the MJO, understanding MJO propagation is expected to have critical 
implications for medium-range weather forecasts and climate predictions 
\cite{Vitart2018}.

The eastward propagation of the MJO is interpreted as a consequence of an 
east-west asymmetry of moisture fields. This asymmetry is intrinsically generated by 
MJO-scale dynamics (equatorial Kelvin-Rossby wave couplet) and climatological 
mean states \cite{Adames2015, Wolding2015, Kim2017, Kang2021}. 
In the Indo-Pacific region, where the background moisture gradient is equatorward 
and eastward, low-level easterlies and cyclonic Rossby gyres associated with 
MJO convection moisten the free troposphere to the east and dry it to the west of 
the large cloud system \cite{Adames2015}, facilitating its eastward movement. 
This widely accepted view forms the basis of representative ``linear" MJO theories 
\cite{Sobel2013, Adames2016a, Wang2016, Wang2017}, which assume only 
MJO-scale perturbations with the time-invariant background states and without 
explicit interactions between the MJO and higher-frequency systems. 
This linearized thinking has been instrumental in evaluating global climate 
model performance \cite{Gonzalez2017, Wang2018b} and interpreting 
future climate projections \cite{Adames2017a, Rushley2019}. However, 
they cannot fully explain key observed features, such as the Maritime Continent 
(MC) barrier effect on MJO propagation \cite{Zhang2017a, Demott2018} and 
the multi-scale structure \cite{Nakazawa1988, Kikuchi2010}, which are critical for 
reliable seasonal predictions and future projections of MJO-related severe 
weather.

Here, using huge-ensemble ($O(10^3)$ members) numerical 
simulations per a single MJO event, we demonstrate that MJO propagation 
cannot be described simply by refining theories under the linearized framework; 
it is more chaotic than previously recognized, originating from cross-scale nonlinear 
interactions. This notion is underpinned by two notable ingredients: (i) multiple regimes 
of MJO propagation emerge under a single background state, depending on 
climatological sea surface temperature (SST) as an external bifurcation parameter; 
and (ii) the selection of the bifurcated regime is governed by subtle differences in 
the cross-scale interactions between high-frequency tropical and extratropical waves 
and the MJO. Revealing these chaotic features, which are not observable if the MJO is 
assumed to be a linear system, together with physical processes is first enabled by 
the exascale supercomputer ``Fugaku". Recent advancements in computing power 
have facilitated an increase in high-resolution ensemble members to reveal attractors 
of subseasonal to seasonal variability with numerous degrees of freedom, as well as 
$O(1$--$10)$-km-scale simulations extending to climate time scales for constructing 
``Digital Earths" \cite{Stevens2024}. The present study relies on the feasibility of 
the former strategy, leveraging the success of global kilometer-scale modeling for 
realistic medium-range weather hindcasts \cite{Miura2007a, Miyakawa2014, 
Stevens2019}.


\section*{Results}
\subsection*{Multiple regimes of the MJO propagation} 
With the 14-km mesh Nonhydrostatic Icosahedral Atmospheric Model (NICAM) 
\cite{Tomita2004, Satoh2014}, we conduct $O(10^3)$-member ensemble hindcast 
simulations for two MJO events realized in early November and December 2018 
(hereafter referred to as Nov-MJO and Dec-MJO). These events are chosen to 
assess how two distinct background states---before and after the onset of the 
Australian monsoon---impact the MJO, with climatologically varying SSTs and 
land surface temperatures during boreal winter, MJO's most active season 
\cite{Zhang2004}. Simulations are initialized with 100 different atmospheric states 
at 00 UTC for each day of a 10-day period, during which MJO convection 
begins forming over the Indian Ocean (Fig. 1, A and B). All simulations span 
45 days, which are sufficiently long to capture the observed eastward migration 
of the large-scale cloud systems to the WP (around 150$^\circ$E). This migration 
occurred around November 20 for Nov-MJO (Fig. 1A) and December 23 for 
Dec-MJO (Fig. 1B).

\begin{figure}
 \centering
 \includegraphics[width=140mm]{./Figure1.eps}
 \caption{{\bf Two MJOs and their regimes revealed by huge-ensemble 
 hindcast simulations.}
 (A and B) Time-longitude diagrams of observed equatorial 
 (10$^\circ$S--10$^\circ$N) outgoing longwave radiation (OLR) at 
 the top-of-atmosphere (shading) and off-equatorial (0$^\circ$--10$^\circ$N) 
 850-hPa meridional winds filtered for 
 westward zonal wavenumbers 1--20 and periods of 5--30 days (contours) for 
 Nov-MJO (A) and Dec-MJO (B). Contour interval is 1.6 m s$^{-1}$, with zero 
 values or lower omitted and areas to the west of 120$^\circ$E masked for visibility.
 (C and D) Number distributions by dates of propagation into the WP for 
 Nov-MJO (C) and Dec-MJO (D). Dark (light) grays indicate the cases 
 of using only 500 (100) members picked up randomly. Bins separated by 
 colored vertical lines belong to the three regimes. 
 (E to G) Time-longitude diagrams of equatorial (10$^\circ$S--15$^\circ$N) 
 column water vapor (shading) and OLR (black) and off-equatorial 
 (0$^\circ$--10$^\circ$N) meridional wind (magenta) anomalies  
 composited for the Nov-1 (E), Dec-1 (F), and Dec-2 (G) regimes. Black 
 (magenta) contour interval is 7 W m$^{-2}$ and 0.6 m s$^{-1}$, with zero 
 values or higher (lower) omitted. Stippling indicates statistical significance of 
 shaded values at the 99\% level.
 } 
\end{figure} 

This experimental design can generate a wide solution space for the MJO propagation 
under given boundary conditions. Figs. 1C and 1D show the frequency distributions 
of the timing when simulated Nov-MJO and Dec-MJO convection reaches the WP, 
respectively. For Nov-MJO, the distribution has a single peak at November 20, when 
the observed propagation is also completed, albeit with a later-phase tail (Fig. 1A). 
Namely, the regime supporting Nov-MJO propagation is unique (denoted as Nov-1 
regime; Fig. 1C), and this regime is chosen with high probability but with its realization 
probabilistic in the model. In contrast, the timing of simulated Dec-MJO propagation 
bifurcates into two distinct windows: around December 22 and 30 (denoted as Dec-1 
and Dec-2 regimes; Fig. 1D). This bimodal distribution is not obtained if the ensemble 
size is limited to 100 or 500 members (Fig. 1D). This result reveals that Dec-MJO 
propagation has the two regimes inherently under a single atmospheric background 
state prescribed by climatologically varying SSTs---analogous to the bifurcation of 
solutions in nonlinear dynamical systems \cite{Strogatz2018}. We can interpret 
the observed Dec-MJO propagation as a realization of the Dec-1 regime (in terms 
of the propagation timing; Fig. 1B).

We create composites of ensemble members for the Nov-1, Dec-1, and Dec-2 regimes 
to examine their MJO propagation in detail. In the Nov-1 and Dec-1 regimes (Fig. 1, E 
and F), MJO-scale convective envelopes migrate eastward to 150$^\circ$E, facilitated 
by westward intrusion of moisture coupled with off-equatorial 850-hPa meridional wind 
variations of synoptic-scale equatorial Rossby waves from the Pacific (fig. S2).
This behavior is consistent with the observed Nov-MJO and Dec-MJO (Fig. 1, A 
and B), confirming the reliability of the simulations, although the MJO in the Dec-1 regime 
exhibits slightly faster propagation and longer stagnation near 150$^\circ$E. In the Dec-2 
regime (Fig. 1G), the moisture-laden westward-propagating wave decays; however,  
moisture accumulation over the WP occurrs through a different process than in 
the Dec-1 regime, offering an alternative MJO-propagation pathway.

\subsection*{Processes supporting the bifurcated regimes of the MJO propagation}

\begin{figure}
 \centering
 \includegraphics[width=140mm, angle=-90]{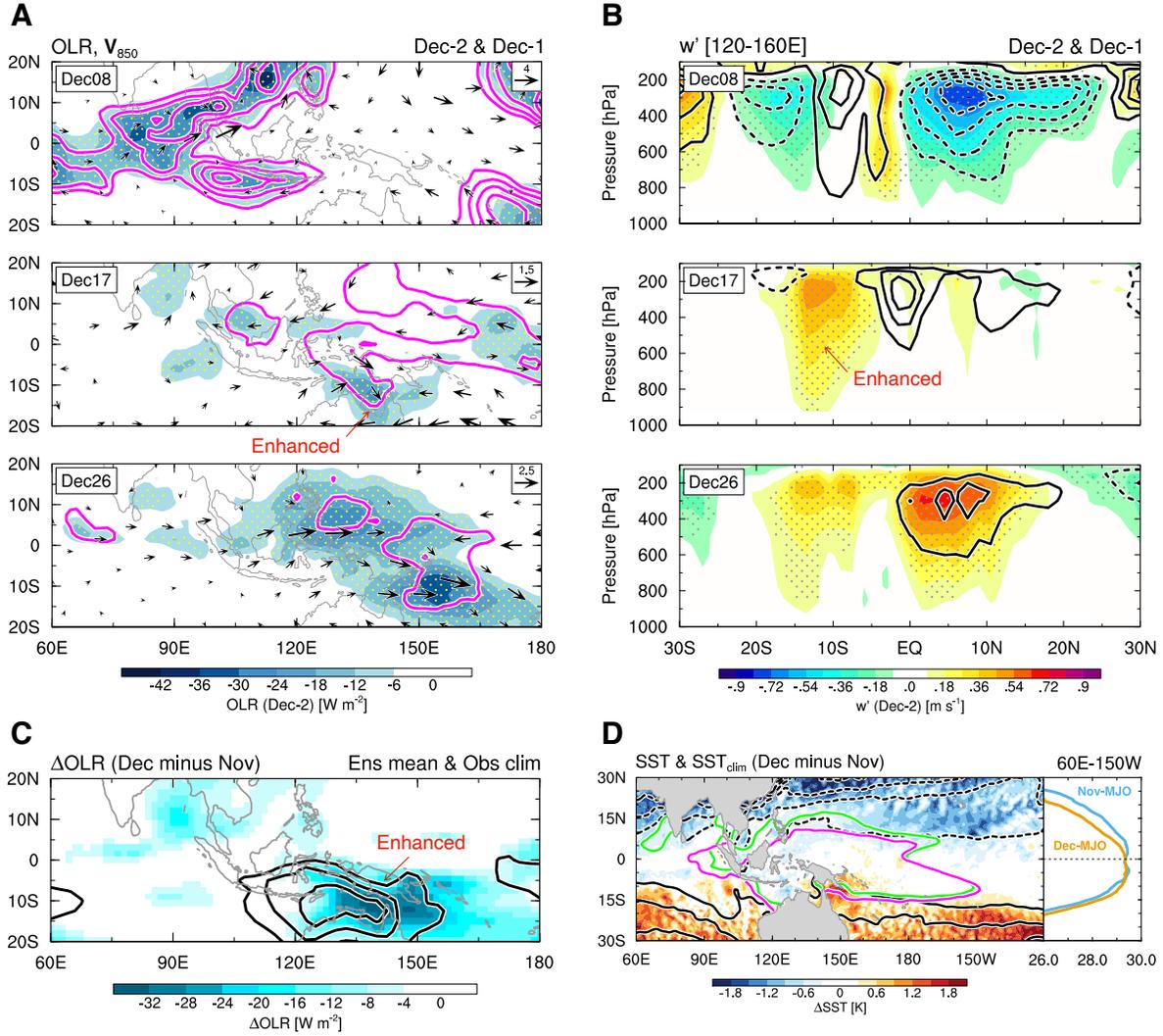}
 \caption{{\bf Time evolution of the MJO propagation regimes in December and 
 background effects on it.}
 (A) OLR (shading) and 850-hPa wind (vectors) anomalies composited for the Dec-2 
 regime on December 8 (top), 17 (middle), and 26 (bottom). 
 Contours indicate OLR anomalies for the Dec-1 regime. Contour interval 
 is 12 W m$^{-2}$, with zero values or higher omitted. Stippling indicates statistical 
 significance of color-shaded values at the 99\% level.
 (B) As in (A), but for latitude-height cross sections of vertical wind anomalies 
 averaged in 120$^\circ$--160$^\circ$E. Contour interval is 0.18 m s$^{-1}$, with 
 zero (negative) values omitted (dashed).
 (C) Differences ($\Delta$: Dec-MJO minus Nov-MJO) in ensemble and simulation 
 period mean of simulated OLR (shading), and those in simulation period mean 
 of observed climatological OLR (contours). Contour interval 
 is 9 W m$^{-2}$, with zero values or higher omitted. 
 (D) Observed raw $\Delta$SST (shading) and climatological $\Delta$SST (black 
 contours), averaged over the simulation periods. Contour interval is 0.5 K, with 
 zero (negative) values omitted (dashed). Green (Magenta) lines indicate the 
 isotherm of 28.5$^\circ$C in time-mean raw SSTs for Nov-MJO (Dec-MJO). 
 Meridional time-mean SST distributions averaged in 60$^\circ$E--150$^\circ$W 
 are plotted in the right panel. 
 }
\end{figure}

We first address why the bifurcation of solutions (i.e., multiple regimes) emerges 
in Dec-MJO propagation. We begin by examining the processes supporting 
the Dec-2 regime, which is the unobserved one. Figure 2A shows the evolution of 
spatial distributions of outgoing longwave radiation (OLR) and 850-hPa wind anomalies 
composited for the Dec-2 regime, compared to the similar composites for Dec-1 MJO. 
While the convective distributions are similar on December 8, they diverge afterwards; 
Dec-2 MJO convection temporarily weakens over the MC on December 17, and then 
reorganizes over the WP on December 26, lagging about one week behind the Dec-1 
regime (cf. Fig. 1D). This reorganization of convection coincided with enhanced 
surface heat fluxes driven by counterclockwise circulation anomalies (10$^\circ$S to 
10$^\circ$N, 120$^\circ$E to 150$^\circ$E), directing winds into equatorially 
asymmetric active convection in the southeastern MC (Fig. 2A for December 17 
and fig. S3).

The equatorially asymmetric convective enhancement is specific to the Dec-2 regime. 
Over the eastern MC, upward motion becomes stronger in 15$^\circ$S to 5$^\circ$S 
on December 17, preceding the realization of large-scale ascent north of the equator 
associated with Dec-2 MJO propagation on December 26 (Fig. 2B). This situation 
indicates that, even though convection near the equator is inactive, meridional 
overturning that balances equatorial radiative cooling can be activated if off-equatorial 
regions favor deep convection. In contrast, equatorial convection remains active on 
December 17 for the Dec-1 regime, supported by the moist wave intrusion (cf. Fig. 1F).

We conjecture that the off-equatorial active convection allowed in the Dec-2 regime 
is linked to seasonal SST changes from November to December.
Compared to Nov-MJO simulations, Dec-MJO simulations exhibit enhanced 
ensemble-mean convective activities in the southeastern MC (Fig. 2C), where the 
equatorial asymmetry of convection is prominent for the Dec-2 MJO propagation 
(Fig. 2, A and B), and this region is marked by climatologically active 
convection during December (Fig. 2C). This corresponds to pronounced SST warming 
in the Southern Hemisphere, including the region of the active convection, 
due to the climatological change in December (Fig. 2D). This southward shift of 
the warm-pool centroid breaks the equatorial symmetry of SSTs, allowing 
the low-level southward cross-equatorial flow to enhance convection.


\subsection*{Climatological SSTs as a key to the MJO propagation regime bifurcation}
From the above results, we hypothesize that the seasonal-mean SST controls 
the presence or absence of the bifurcating regimes of the MJO propagation. 
Here, we verify this hypothesis using additional huge-ensemble simulations. 
We first conduct a 1,000-member sensitivity experiment for Dec-MJO, 
where the oceanic boundary condition is replaced with that used in the Nov-MJO 
simulation. Under the November-SST condition, the Dec-2 regime does not emerge 
and the timing of Dec-MJO propagation is almost uniquely determined, although 
the timing of the realization of this unique regime is slightly earlier than the Dec-1 
regime (Fig. 3A). As additional support, we also conduct a 1,000-member sensitivity 
experiment for Nov-MJO using SSTs from the Dec-MJO simulation. The timing 
of Nov-MJO propagation becomes more diverse than the control experiment with 
the unique regime (Fig. 3, B and C), and the propagation at a later timing (around 
November 28) is possible especially for the simulation with long lead time (Fig. 3C). 
This indicates that the later regime emerges under strong influences of the ocean. 
These results confirm that the bifurcation of the MJO propagation regimes originates 
from the equatorial asymmetry installed in the climatological SSTs.


\begin{figure}
 \centering
 \includegraphics[width=40mm, angle=-90]{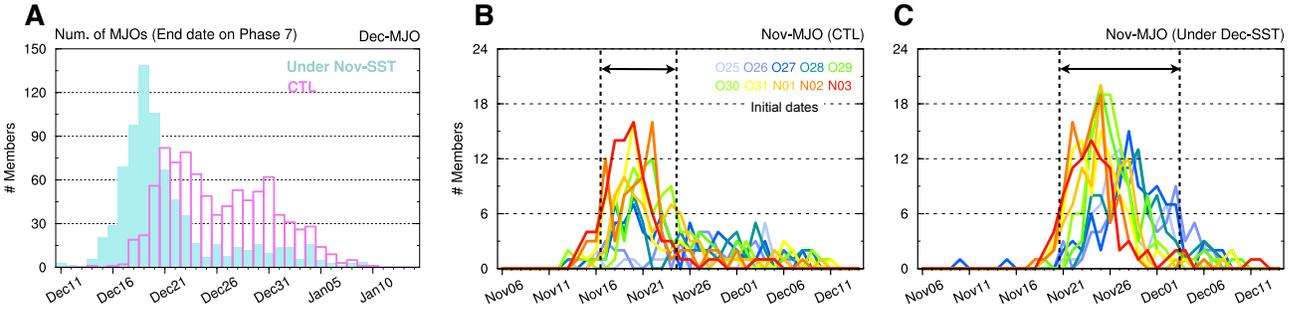}
 \caption{{\bf Impacts of seasonal SST components on the timing of MJO propagation.}
 (A) Number distributions by dates of propagation into the WP for the Dec-MJO 
 simulation under SSTs used in the Nov-MJO simulation (filled) and for the control 
 Dec-MJO simulation (open). 
 (B and C) As in (A), but for the control Nov-MJO simulation (B) and the Nov-MJO 
 simulation under SSTs used in the Dec-MJO simulation (C). Results for each 100 
 members with 10 different initial dates are plotted by different colors.
 }
\end{figure}

\subsection*{Tropical-extratropical interaction as a cause of the regime selection}

\begin{figure}
 \centering
 \includegraphics[width=140mm, angle=-90]{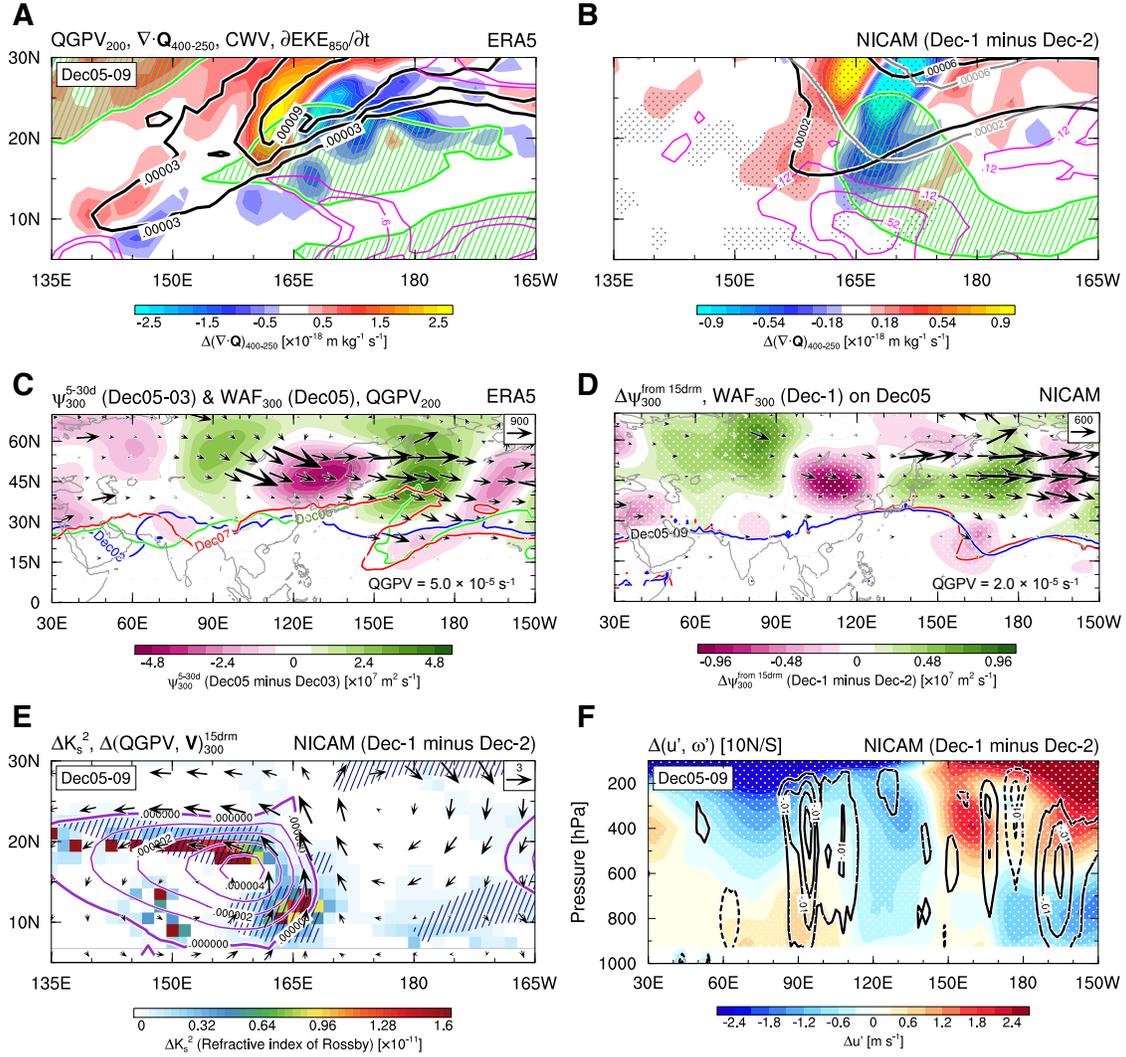} 
 \caption{{\bf Regime selection by the MJO-extratropical interaction.}
 (A) Observed Q-vector divergence in 400-250 hPa (shading), 200-hPa 
 QGPV (black), and column water vapor (CWV) anomalies ($>$ 2.5 kg m$^{-2}$; 
 hatched) averaged over December 5--9, and 850-hPa EKE tendency on 
 December 7 (magenta).
 (B) As in (A), but for the simulation. QGPV for the Dec-1 (black) and Dec-2 (gray) 
 regimes, and differences ($\Delta_{\rm{Dec}}$; Dec-1 minus Dec-2 regime) in the 
 other variables are plotted. Stippling indicates statistical significance of shaded 
 values at the 99\% level (Same below).
 (C) Observed 300-hPa Takaya-Nakamura wave activity flux \cite{Takaya2001} 
 on December 5 (vectors), 200-hPa QGPV during December 3--7 (colored contours), 
 and differences (December 5 minus 3) in 5--30-day bandpass-filtered 300-hPa 
 stream function ($\psi$; shading).
 (D) As in (C), but for the simulation. Shading displays $\Delta_{\rm{Dec}}\psi$ 
 on December 5. Red (blue) contours indicate time-mean 200-hPa QGPV for 
 the Dec-1 (Dec-2) regime.
 (E) Time-mean $\Delta_{\rm{Dec}}$ in the refractive index of Rossby waves 
 (shading), 15-day running mean 300-hPa QGPV (contours), winds (vectors), 
 and QGPV gradients ($>$ 2.0 $\times$ 10$^{-12}$ s$^{-1}$ m$^{-1}$; hatched).
  (F) Time-mean $\Delta_{\rm{Dec}}$ in equatorial (10$^\circ$S--10$^\circ$N) 
 zonal wind (shading) and vertical $p$-velocity (contours) anomalies.
 }
\end{figure}

As a mechanism explaining the selection of the bifurcated regimes for Dec-MJO 
propagation, we find a difference in the strength of the tropical-extratropical interaction 
between the two regimes. In the observation, an extratropical trough intrudes into 
the tropical WP in the upper troposphere prior to the MJO propagation 
(December 5 to 9). This trough forces large-scale ascent and moistens 
the troposphere to the east (Fig. 4A), creating conditions conductive to 
the development and maintenance of the westward-propagating equatorial 
Rossby wave that leads Dec-MJO propagation (Fig. 1B). This interpretation is 
supported by the positive tendency of westward-propagating synoptic-scale 
eddy kinetic energy (EKE) at 850 hPa in 160$^\circ$E to 180$^\circ$ and its 
budget analysis (Fig. 4A and fig. S4).

The role of the tropical-extratropical interaction in the regime selection is confirmed 
by the huge-ensemble simulations. The Dec-1 regime, which follows the observed 
processes of Dec-MJO propagation (Fig. 1F), involves the trough intrusion from 
the extratropics (Fig. 4B), consistent with the observations. Meanwhile, this signal is 
weaker and more eastward-displaced in the Dec-2 regime. Consequently, the Dec-1 
regime features a more robust large-scale ascent dynamically forced by the extratropical 
influences, along with a moister environment in the off-equatorial region. These factors 
lead to a stronger positive tendency of synoptic-scale EKE. This contrast delineates 
the watershed between the two regimes: the strong (weak) extratropical influence 
fosters (inhibits) the development of the westward-propagating equatorial Rossby 
wave before Dec-MJO propagation, resulting in the Dec-1 (Dec-2) regime.

The trough intrusion from the extratropics is a highly transient process, realized as 
the refraction of extratropical Rossby waves. An observed subtropical jet, 
measured by quasi-geostrophic potential vorticity (QGPV) at 200 hPa, meanders 
in 150$^\circ$E to 180$^\circ$ suddenly on December 5 (Fig. 4C). Simultaneously, 
mid-latitude transient Rossby waves disperse energy into the tropics in 
150$^\circ$--170$^\circ$E. This energy dispersion is active in the Dec-1 regime, 
sustaining stronger cyclonic circulation and more evident trough intrusion in 
10$^\circ$--30$^\circ$N, 150$^\circ$--170$^\circ$E than the Dec-2 regime (Fig. 4D). 
In the Dec-1 regime, the higher refractive index of Rossby waves supports 
the clearer refraction of the extratropical Rossby waves, under enhanced 
background QGPV gradients linked to stronger cyclonic circulation in 
the upper troposphere (Fig. 4E).

The upper-level background cyclonic circulation that influences the extratropical 
Rossby waves appears to be generated as a Matsuno-Gill response to the MJO 
convection \cite{Hendon1994a}, which is active over the western MC at this time. 
This inspection raises the hypothesis that the intensity of the MJO-related circulations 
self-regulates the degree of the extratropical-tropical interaction, verified by our 
simulations. Figure 4F shows the differences in the equatorial zonal circulation 
before the MJO propagation between the two regimes. As expected, the Dec-1 regime 
exhibits stronger ascent associated with MJO convection in 90$^\circ$--120$^\circ$E, 
accompanied by more predominant upper-level westerlies over the WP. 
These features constitute the stronger cyclonic circulation in Fig. 4E. To sum up 
the above results, the differences in the two-way interaction between the extratropical 
Rossby waves and MJO-scale tropical circulation are responsible for splitting 
Dec-MJO propagation into the Dec-1 and Dec-2 regimes.

\section*{Discussion}
If the dynamics of the MJO were essentially linear, as assumed by previous 
theoretical studies \cite{Sobel2013, Adames2016a, Wang2016, Wang2017}, 
the bifurcation in the timing of MJO propagation that depends on the seasonal 
mean SST distributions could never been seen. Our huge-ensemble simulations 
prove that the MJO dynamics cannot be explained by a linear theory. Instead, 
the MJO should be regarded as a nonlinear mode that allows multiple regimes and 
involves the interactions between the MJO, synoptic-scale equatorial waves, and 
extratropical disturbances. The physical mechanism we identify as causing 
chaotic MJO propagation advances the insight from Chen (2024) \cite{Chen2024}, 
who showed the chaos in the evolution of the MJO amplitude using a Koopman 
model but left its physical origin a open question. The key physics we found aligns 
with features reported in the literature.
Westward-propagating equatorial Rossby waves have been observed to interfere 
with MJO convective envelopes \cite{Roundy2004, Masunaga2006}. 
Regarding extratropical impacts, case studies of other MJO events 
suggested the possibility that MJO propagation is occasionally facilitated by 
extratropical wave-breaking-induced convection over the tropical central Pacific 
\cite{Meehl1996, Moore2010}, as in the Dec-1 regime. Furthermore, a statistical 
analysis of observations showed that a stronger cross-equatorial local Hadley 
circulation around the MC favors MJO propagation \cite{Takasuka2025}, implying 
the robustness of the existence of the Dec-2 regime realized by the equatorially 
asymmetric convective activities (Fig. 2). These reports collectively support 
the need to reconsider the existing MJO theories in light of nonlinearity. 

Our findings are applicable to interpreting the seasonality of the MJO activity. 
It is well known that the MJO is most active from December to February (DJF) 
\cite{Zhang2004}. This seasonality can be explained by the existence of multiple 
regimes of MJO propagation under the equatorially asymmetric SSTs, which are 
prominent in DJF. Supportive evidence for this idea is that over 60\% of MJO 
events in November fail to propagate into the WP, whereas only about 25\% of 
events fail to propagate in DJF \cite{Kerns2020}. The higher likelihood of MJO 
propagation after December is attributable to the greater variety of regimes 
with different moistening processes over the WP, compared to November, 
as revealed by our simulations. To further elucidate the relationship between 
the seasonality of MJO activity and the number of MJO propagation regimes, 
it would be valuable to categorize MJO-propagation features into cases involving 
the interactions with synoptic-scale tropical and extratropical dynamics 
(like the Nov-1 and Dec-1 regimes) and more large-scale forced cases 
(like the Dec-2 regime).

Our results also have implications for the prediction of MJO propagation, 
which is important to subseasonal to seasonal weather forecasts \cite{Vitart2018}, 
including those utilizing machine learning (ML). At the beginning of boreal winter 
(e.g., November), whether the MJO can propagate into the WP appears uncertain, 
as our simulation suggests that the realization of the Nov-1 regime is probabilistic. 
In contrast, MJO events in the middle of boreal winter (i.e., DJF) 
exhibit greater uncertainty regarding the timing of MJO propagation rather than 
the completion of that. This suggests that predicting both the timing and occurrence 
of MJO propagation during DJF remains inherently challenging, even if prediction 
models have the ability to simulate the MJO. This highly probabilistic nature of 
MJO propagation implies limited prediction skill using ML models trained solely on 
observational data \cite{Martin2022} and/or conventional global climate model (GCM) 
data \cite{Shin2024, Suematsu2022b}. Observational data alone cannot encompass 
the full probability space of the MJO, and conventional GCMs struggle to accurately 
simulate synoptic-scale variations affecting MJO dynamics \cite{Bartana2023}. 
Better AI-based prediction of the MJO may be achieved by training models on 
huge-ensemble storm-resolving simulations that capture the bifurcating solutions 
of MJO propagation arising fromthe cross-scale interactions.

In conclusion, nonlinearity, a necessary condition for chaos, is involved in MJO 
propagation through moisture advection by synoptic-scale waves. Also, the bifurcated 
two regimes, which emerge in response to changes of the equatorial asymmetry of 
SSTs, transition non-periodically via the tropical-extratropical interaction. These findings 
show that MJO propagation can be viewed as a deterministic chaotic phenomenon, and 
that its more skillful prediction necessitates focus on the probabilistic space revealed by 
huge-ensemble simulations.

\section*{Materials and Methods}
\subsection*{Model and simulation setups}
In this study, we use the Nonhydrostatic 
Icosahedral Atmospheric Model (NICAM) \cite{Tomita2004, Satoh2014}.
A set of fully compressible three-dimensional nonhydrostatic dynamical equations 
are used, and their discretization is on an icosahedral A-grid system with 
spring dynamics on the sphere \cite{Tomita2002}. A globally quasi-uniform horizontal 
grid interval we adopted is about 14 km. This resolution can represent multi-scale 
structure emerging from cloud systems explicitly and globally, including the MJO 
\cite{Miyakawa2014, Takasuka2024}, although it is still far from a convection-resolving 
scale. Vertical layers are 38, extending to an altitude of about 40 km above 
the sea surface. The physics schemes are the same as the configuration used in 
the ``MJO run" in Takasuka et al. (2024) \cite{Takasuka2024}. 

The atmospheric states in the 1,000-member ensemble 45-day simulations for 
the Nov-MJO and Dec-MJO events are initialized at 00UTC each day during a 10-day 
period shown in Figs. 1A and 1B. This 10-day period for 
Nov-MJO and Dec-MJO is October 25--November 3 and November 23--December 2, 2018, 
respectively. The atmospheric initial conditions are obtained from NICAM-Local Ensemble 
Transform Kalman Filter (LETKF) Japan Aerospace Exploration Agency (JAXA) Research 
Analysis (NEXRA) data set \cite{Kotsuki2019}, which has 100 ensemble members every 
6 hour with a horizontal resolution of 1.25$^\circ$ $\times$ 1.25$^\circ$. 
We use all the 100 members of NEXRA at 00UTC during the above periods (10 days 
$\times$ 2), and thus we can create the two sets of the 1,000-member simulation as 
the control hindcast experiments for Nov-MJO and Dec-MJO. The oceanic state is 
initialized by the National Oceanic and Atmospheric Administration (NOAA) Optimum 
Interpolation Sea Surface Temperature (OISST) Version 2.1 \cite{Reynolds2007}, and 
the oceanic boundary conditions of SST and sea ice fraction are given by the linear 
temporal interpolation of the weekly OISST data. The land initial condition is the monthly 
mean climatology of the NICAM simulation with a 220-km horizontal mesh. 
These oceanic and land data are the same for all the ensemble simulations.

In addition to the control hindcast experiments, 
we conduct the two sets of the 1,000-member ensemble sensitivity experiments to 
examine the impacts of SST distributions on the behavior of solutions in the MJO propagation. 
In one experiment, the oceanic initial and boundary conditions used in the Dec-MJO 
control simulations are replaced with those from the Nov-MJO control simulation. 
Specifically, this 1,000-member simulation spans the calendar period from November 23, 
2018 to January 16, 2019, while the oceanic conditions are taken from October 25 to 
December 18, 2018. In the other experiment, this relationship between the calendar 
period and oceanic conditions is reversed. All the configurations other than the oceanic 
conditions are identical to those in the control simulations.


\subsection*{Observational data}
We use the three observational data sets. First, interpolated daily OLR obtained from 
the NOAA polar-orbiting satellite \cite{Liebmann1996} is used as a proxy for deep convective 
activities. The horizontal resolution is 2.5$^\circ$ $\times$ 2.5$^\circ$. The data cover the period 
from January 1979 through December 2019, used for constructing phase space that 
describes the time evolution of MJO convection (described later) as well as for checking 
the observed Nov-MJO and Dec-MJO propagation. We also use 6-hourly snapshots 
from the ERA5 reanalysis data \cite{Hersbach2020} with a horizontal resolution of 
1.5$^\circ$ $\times$ 1.5$^\circ$, averaged to daily values that span the period from 
January 1979 to December 2022. We depict the three-dimensional variables of zonal 
and meridional winds ($u$ and $v$), vertical p-velocity ($\omega$), temperature ($T$), 
specific humidity ($q_v$), and geopotential ($\Phi$). The three-dimensional data used 
for the analyses have 15 pressure levels, spanning from 1,000 to 50 hPa. 
In addition, we take the two-dimensional surface pressure data to calculate column-integrated 
water vapor content from the surface to 100 hPa. Daily anomalies are calculated by 
the removal of the first three harmonics of daily climatologies for 1979--2022, before we apply 
spatial and/or temporal filtering to any variables. Lastly, we use the observed daily SST 
data from OISST Version 2.1 \cite{Reynolds2007}. The daily climatology used in Fig. 2D 
is calculated from the 1971--2000 base period. The horizontal resolution is 0.25$^\circ$ 
$\times$ 0.25$^\circ$.

\subsection*{Identifying the MJO propagation and its composite}
To identify the completion of the MJO propagation in our ensemble simulations, we monitor 
the time evolution of the MJO index introduced by Kikuchi et al. (2012) \cite{Kikuchi2012}. 
The MJO index is calculated for every ensemble member. Originally, the MJO index is 
constructed by two leading principal components (PC1 and PC2) obtained from 
the extended empirical orthogonal function (EEOF) analysis for the long-term daily 
OLR data in the tropics. The period of our simulations, however, is only 45 days, which 
are not sufficiently long to calculate the statistically robust EEOFs. Thus, we adopt the 
hybrid usage of the simulation and observational data to obtain the MJO index here. 
This treatment has been taken similarly in a previous study \cite{Shibuya2021}. 
The specific procedure is as follows:

\begin{enumerate}
 \item We calculate two leading extended empirical orthogonal functions (EEOF1 and 
 EEOF2) of observed intraseasonal OLR anomalies in the tropics (30$^\circ$S--30$^\circ$N, 
 0$^\circ$--360$^\circ$E) from December to February in 1979--2019. Here, intraseasonal 
 OLR anomalies are obtained by applying Lanczos band-pass filter \cite{Duchon1979} 
 with 25--90-day cutoff periods and 201 weights to daily unfiltered OLR anomalies. 
 The daily unfiltered OLR anomalies are calculated by the removal of the first three 
 harmonics of daily climatology for 1979--2019. The EEOF analysis uses three
  time-lagged data: days $-$10, $-$5, and 0.

  \item To calculate the MJO index for every ensemble member, we prepare for 
  the simulation-based daily OLR data projected onto the two observation-based 
  EEOFs (denoted as EEOFs$^{\it Obs}$). We first subtract the time mean of the simulated 
  OLR data during the entire period of the simulation for each member (denoted as 
  OLRA$^{\it Sim}$). If OLRA$^{\it Sim}$ are projected onto EEOFs$^{\it Obs}$ with 
  the time lags (days $-$10, $-$5, and 0), we cannot obtain the MJO index on 
  the first 10 simulation days. Thus, the 10-day time series of observed OLR 
  anomalies (denoted as OLRA$^{\it Obs}$) are appended before the beginning of 
  the time series of OLRA$^{\it Sim}$. For instance, for a 45-day time series of 
  OLRA$^{\it Sim}$ that starts at $t=t_0$ [day], OLRA$^{\it Obs}$ from $t= t_0-10$ to 
  $t=t_0-1$ are attached. Here, OLRA$^{\it Obs}$ are defined as the deviations 
  from the time mean during the simulation period (i.e., $t_0 \le t \le t_0+45$) 
  for the consistency of the base period of the anomalies with OLRA$^{\it Sim}$.
  
  \item We project the 55-day OLR data created above onto the two EEOFs$^{\it Obs}$, 
  yielding the 45-day time series of PC1 and PC2. Then, we apply 5-day running mean to 
  these two PCs to remove the high-frequency noisiness, and define the amplitudes 
  ($A=\sqrt{PC1^2+PC2^2}$) and phases ($\alpha=\tan^{-1}(PC1/PC2)$) of the MJO 
  index.
  
\end{enumerate}

Next, we track the day-to-day evolution of the amplitudes and phases of the MJO index 
to identify when the MJO convection completes its propagation into the WP. 
The phases are divided into eight (Phases 1 to 8) by $\pi/4$ phase angle 
(see fig. S1; the consecutive phase progression captures the eastward propagation 
of large-scale convective envelopes successfully). We impose four criteria on 
the tracking of the MJO index.

\begin{itemize}
\item Passing through Phase 1 within the first 20 days.
\item $A > 0.4$ during tracking, except that the number of days with $A \le 0.4$ is 
4 days or less.
\item No more than one phase skipping, and no more than three phase recession. 
\item Completion of tracking up to Phase 7 or 8 (i.e.. from Phase 6 to 7, 5 to 7, 
and 6 to 8), or that of staying at Phase 6 for more than 7 days with all the above 
conditions satisfied. 
\end{itemize}

The first criterion ensures the robust MJO initiation over the Indian Ocean (IO); 
the second and third ones require the eastward migration of the MJO structure with 
certain strength; and the fourth criterion confirms the MJO propagation into the WP. 
The timing of the MJO propagation into the WP (cf. Fig. 1, C and D) is specified 
as the day on which the amplitudes take the maximum, of those that meet 
the fourth criterion.

After identifying the MJO propagation timing, we apply the composite 
analyses to the ensemble members categorized into the three representative 
regimes of the MJO propagation: the Nov-1, Dec-1, and Dec-2 regimes. 
Unless otherwise noted, the anomalies of any variables used to create 
composites are defined as deviations from the simulation period mean 
for each ensemble member. Statistical significance of the composite anomalies 
and their differences between the regimes is assessed by a two-tailed 
Student's t test, on the assumption that the different time evolutions among 
the ensemble members are statistically independent. 

\subsection*{Moist static energy budget}
We conduct the column-integrated moist static energy (MSE) budget analysis to 
understand the moistening processes responsible for the MJO propagation into the WP. 
This relies on the fact that the MSE tendency explains tropical moisture variations 
well because the horizontal temperature gradient is weak in the tropics. 
The budget equation used here is 
\begin{equation}
\langle \partial_t h \rangle = -\langle \mathbf{v}_h \cdot \nabla h \rangle 
- \langle \omega\partial_p h \rangle + \mathit{SHF} + \langle Q_R \rangle
\end{equation}
where $h=C_pT+\Phi+L_v q_v$ is MSE ($C_p = 1004.69$ J kg$^{-1}$ is the specific 
heat at constant pressure; and $L_v = 2.50084$ $\times$ 10$^6$ J kg$^{-1}$ is 
the latent heat for vaporization); $\mathbf{v}_h$ is the horizontal velocity vector; 
$SHF$ is the sum of surface latent and sensible heat fluxes; and $\langle Q_R\rangle$ 
is radiative heating. The angle brackets indicate mass-weighted vertical integration 
from the surface to 100 hPa. Note that $\langle Q_R\rangle$ is calculated as 
the difference of radiative fluxes between the top of the atmosphere and 
the surface. The budget terms are evaluated with 6-hourly values, and 
then they are averaged to the daily values.

\subsection*{Eddy kinetic energy budget}
The eddy kinetic energy (EKE) budget is analyzed to explain processes that affect 
the equatorial Rossby-wave activities involved in the MJO propagation. The budget 
equation is
\begin{equation}
\frac{\partial \overline{K'}}{\partial t} =
- \underbrace{\overline{\mathbf{v}_h' (\mathbf{v}' \cdot \nabla) \mathbf{v}_h}}_{\mathit{KmKe}}
- \underbrace{\overline{\mathbf{v}} \cdot \nabla \overline{K'}}_{\mathit{AmKe}}
- \underbrace{\overline{\mathbf{v}'} \cdot \nabla \overline{K'}}_{\mathit{AeKe}}
- \underbrace{\frac{R}{p} \overline{\omega' T'}}_{\mathit{PeKe}}
- \underbrace{\overline{\nabla \cdot (\mathbf{v}' \Phi')}}_{\mathit{GKe}}
+ (\mathit{Residual})
\end{equation}
where $K' = (u'^2 + v'^2)/2$ is EKE; $\mathbf{v}$ is the three-dimensional wind vector; 
and $R =  287.05$ J kg$^{-1}$ is the dry gas constant. The terms on the right-hand 
side are physically explained as follows: $KmKe$ is the barotropic conversion from 
mean flows to EKE; $AmKe$ and $AeKe$ are the EKE advection by mean and 
eddy flows, respectively; $PeKe$ is the baroclinic conversion from the eddy available 
potential energy; $GKe$ is the EKE dispersion via the work done by pressure 
gradient forces; and Residual includes diffusive processes. $KmKe$, $PeKe$, 
and $Residual$ correspond to source and sink terms, and the others contribute 
to the redistribution of EKE.

For the observations (simulations), primes and overbars denote values filtered for 
zonal wavenumbers $-$20 to $-$1 and periods of 5--30 days (deviations from 
11-day running mean) and 11-day running mean, respectively. The definition of 
the primes differs between the observations and simulations because the simulation 
data period is too short for filtering. However, this is not an issue, as both definitions 
successfully capture variations associated with the equatorial Rossby waves 
(Fig. 1, B and E to F). To prevent the loss of the simulated data at their edges when 
calculating the running mean, we append the observational data before and after 
the simulation periods prior to the budget calculation. For filtering, we use fast 
Fourier transforms in space and a 201-point Lanczos band-pass filter 
\cite{Duchon1979} in time.

\subsection*{Diagnoses based on quasi-geostrophic dynamics}
We provide some diagnoses based on quasi-geostrophic (QG) dynamics to understand 
the tropical-extratropical interaction as a cause of the regime selection of the Dec-MJO 
propagation. Here, we describe the methodologies used to calculate the four quantities: 
QG potential vorticity (QGPV), Q-vector, the phase-independent wave activity flux, and 
the refractive index of Rossby waves. In this study, we adopt the $\beta$-plane 
approximation with a reference latitude 35$^\circ$N for all the QG diagnoses. Note that 
the wave activity flux and refractive index of Rossby waves are calculated based on 
the composite fields, not on individual ensemble members. This can capture the behavior 
of Rossby waves directly related to the overall features of the regimes of interest.

\paragraph*{Definition of the quasi-geostrophic potential vorticity (QGPV)}
The QGPV on the pressure coordinate is defined as
\begin{equation}
q = f_0 + \beta y + \frac{\partial^2 \psi}{\partial x^2}
+ \frac{\partial^2 \psi}{\partial y^2}
+ \frac{\partial}{\partial p} \left( \frac{f_0^2}{S^2} \frac{\partial \psi}{\partial p} \right)
\end{equation}
where $f_0$ is the Coriolis parameter at the reference latitude (35$^\circ$N); 
$S^2 = -\alpha(\partial\ln\theta/\partial p)$ is the static stability parameter ($\alpha$ 
is the specific volume; and $\theta$ is the potential temperature); and 
$\Psi=(\Phi-\Phi_{\rm ref})/f_0$ is the quasi-geostrophic stream function. We use 
$S^2$ and $\Phi_{\rm ref}$ calculated as their time mean (1979--2022 and 
the simulation period for the observations and simulations, respectively) 
and areal mean over the Northern Hemisphere, so they are the functions 
of only $p$. To remove small-scale variations that do not follow the geostrophic 
motion largely, we apply 9-point spatial smoothing to the obtained QGPV.

\paragraph*{Q-vector analysis}
We calculate the divergence of the Q-vector \cite{Hoskins1978} to diagnose 
the vertical motions forced by the QG dynamics. Horizontal components of 
the Q-vector are represented by 
\begin{equation}
\mathbf{Q} \equiv -\frac{R}{p} \nabla_h \mathbf{v}_{gh} \cdot \nabla_h T
= f_0 \left(
\begin{array}{c}
- \psi_{xy} \psi_{xp} + \psi_{xx} \psi_{yp} \\
- \psi_{yy} \psi_{xp} + \psi_{xy} \psi_{yp}
\end{array}
\right)
\end{equation}
where $\nabla_h$ is the horizontal gradient operator; and $\mathbf{v}_{gh}$ 
is the horizontal velocity vector associated with geostrophic motions. 

\paragraph*{Calculation of the phase-independent wave activity flux}
We use the phase-independent wave activity flux introduced by Takaya and 
Nakamura (2001) \cite{Takaya2001} to evaluate the Rossby-wave energy 
dispersion. We consider the contributions from both stationary and migratory 
QG eddies on a pressure level. The equation of the corresponding wave 
activity flux is as follows:
\begin{equation}
\mathbf{W}_h =
\frac{1}{2 |\overline{\mathbf{v}_h}|}
\left(
\begin{array}{l}
\overline{u}\left( \psi'_x{}^2 - \psi' \psi'_{xx} \right)
+ \overline{v}\left( \psi'_x \psi'_y - \psi' \psi'_{xy} \right) \\[5pt]
\overline{u}\left( \psi'_x \psi'_y - \psi' \psi'_{xy} \right)
+ \overline{v}\left( \psi'_y{}^2 - \psi' \psi'_{yy} \right)
\end{array}
\right)
+ \mathbf{C}_{\overline{\mathbf{v}_h}} M
\end{equation}
Here,  $\mathbf{C}_{\overline{\mathbf{v}_h}} \equiv 
C_P(u/|\mathbf{v}_h|, v/|\mathbf{v}_h|)^T$ is the horizontal vector that 
represents the wave phase propagation projected onto the direction 
of $\overline{\mathbf{v}_h}$ (the horizontal vector of background winds); and 
$M=1/2 (A+E)$ with $A=q'^2/(2|\nabla_h \overline{q}|)$ and 
$E=e/(|\overline{\mathbf{v}_h}| -C_P)$, where $C_P$ is the wave phase 
speed in the direction of $|\overline{\mathbf{v}_h}|$; and $e$ is the wave 
energy, represented as 
\begin{equation}
e = \frac{1}{2} \left[
\left( \frac{\partial \psi'}{\partial x} \right)^2
+ \left( \frac{\partial \psi'}{\partial y} \right)^2
+ \frac{f_0^2}{S^2} \left( \frac{\partial \psi'}{\partial p} \right)^2
\right]
\end{equation}
Note that we apply four times of 9-point spatial smoothing to $q$ and 
$\overline{q}$ when calculating $A$. 

As in the EKE budget analysis, the background states (overbars) and perturbations 
(primes) are defined differently between the observations and simulations because of 
the limitation of the simulated data period. For the observations, the overbars and primes 
denote values filtered for periods of 30 days and more and 5--30 days, respectively. 
Meanwhile, the overbars and primes for the simulated data indicate 15-day running 
mean and deviations from it, respectively. 

To calculate the wave activity flux from migratory QG eddies 
(i.e., $\mathbf{C}_{\overline{\mathbf{v}_h}} M$), we estimate $C_P$ at each grid point, 
following Takaya and Nakamura (2001) \cite{Takaya2001}. First, on the day of interest 
($t_0$), we first compute the two correlation coefficients of 
$\Phi' (t)$ $(t_0-10 \le t \ [{\rm day}] \le t_0+10)$ between a base point and other grids within 
a 40$^\circ$ $\times$ 40$^\circ$ grid box centered at the base point, imposing a lag of 
$-$1 day and 1 day. Then, using the created one-point correlation maps, we trace 
the grid point with the maximum positive correlation from the negative to positive lag 
to estimate the actual wave phase propagation. Lastly, the horizontal vector of 
the actual phase propagation at the local point is projected onto the direction of 
local background winds ($|\overline{\mathbf{v}_h}|$), and then $C_P$ is estimated 
locally as the magnitude of the projected vector.

\paragraph*{Refractive index of Rossby waves}
Following Nishii and Nakamura (2004) \cite{Nishii2004}, we calculate the refractive 
index of Rossby waves as follows:
\begin{equation}
K^2_s =
\frac{|\nabla_h \overline{q}|}{|\overline{\mathbf{v}_h}| - C_P}
- \frac{f_0^2}{4 N^2 H_0^2}
\left(
1 - 4 H_0 N \frac{d N^{-1}}{dz^*}
+ 4 H_0^2 N \frac{d^2 N^{-1}}{d{z^*}^2}
\right)
\end{equation}
where $N$ is the Brunt-Vaisala frequency averaged over the Northern Hemisphere 
and simulation period; $H_0$ is the scale height (set to 8.5 km); and 
$z^*=-H_0 \ln(p/p_0)$ (i.e., log-pressure vertical coordinate). The overbars indicate 
15-day running mean. As in the calculation of the wave activity flux, four times of 
9-point spatial smoothing are applied to the $\overline{q}$ field.

\bibliography{scibib}

\bibliographystyle{Science}

\section*{Acknowledgments}
We would like to thank Hiroaki Tatebe for helpful comments on the design of the sensitivity 
experiments. \\
{\bf Funding}: D.T. was supported by JSPS KAKENHI Grant JP20H05728 and 
JP24K22893. T.S. was supported by JSPS KAKENHI Grant JP21K13991. H.M. was 
supported by JSPS KAKENHI Grant JP23H01243 and JP23K25939. 
D.T. and M.N. were supported by the Ministry of Education, Culture, Sports, Science, 
and Technology (MEXT) as ``Program for promoting 
researches on the supercomputer Fugaku" JPMXP1020200305, Large Ensemble 
Atmospheric and Environmental Prediction for Disaster Prevention and Mitigation. \\
{\bf Author contributions}: D.T. and H.M. conceptualized this work. D.T. performed all 
the numerical experiments and data analysis. He also produced all the figures and wrote 
the original draft. T.S. contributed to the design and interpretation of the sensitivity 
experiments. M.N. led the acquisition of funding and computational resources. 
All authors discussed the results and contributed to the last version of the manuscript. \\
{\bf Competing interests}: The authors have no conflicts of interest in this work. \\
{\bf Data and materials availability}: Interpolated daily outgoing longwave radiation (OLR) 
data obtained from the NOAA polar-orbiting satellite 
are downloaded from \url{https://psl.noaa.gov/data/gridded/data.olrcdr.interp.html}. 
Observational three-dimensional variables of zonal and meridional winds, 
vertical $p$-velocity, temperature, specific humidity, and geopotential are 
from the European Center for Medium range Weather Forecasting (ECMWF) 
ERA5, 
available at 
\url{https://cds.climate.copernicus.eu/datasets/reanalysis-era5-pressure-levels?tab=overview}.
The observed daily SST data from OISST Version 2.1 
are downloaded 
from \url{https://psl.noaa.gov/data/gridded/data.noaa.oisst.v2.highres.html}.
The NICAM-Local Ensemble Transform Kalman Filter (LETKF) Japan Aerospace Exploration 
Agency (JAXA) Research Analysis (NEXRA) data set, used for the atmospheric initialization 
of our huge-ensemble simulations, is available at 
\url{https://www.eorc.jaxa.jp/theme/NEXRA/guide.htm}. The simulation and observational 
data and scripts needed to create the figures are stored at Dryad. 

\section*{Supplementary materials}
Supplementary Text\\
Figs. S1 to S4


\clearpage



\end{document}